\documentclass[sigconf]{acmart}

\AtBeginDocument{%
  \providecommand\BibTeX{{%
    \normalfont B\kern-0.5em{\scshape i\kern-0.25em b}\kern-0.8em\TeX}}}

\setcopyright{acmcopyright}
\copyrightyear{2019}
\acmYear{2019}

\acmConference[Mobiquitous2019]{16th EAI International Conference on Mobile and Ubiquitous Systems: Computing, Networking and Services}{Nov 2019}{Houston, USA}

\usepackage{amsmath,amssymb,amsfonts}
\usepackage{algorithm}
\usepackage{algorithmic}
\usepackage{graphicx}
\usepackage{textcomp}
\usepackage{amsmath}
\usepackage{xcolor}

\begin{document}

\title{Impact of consensus on appendable-block blockchain for IoT}


\author{Roben C. Lunardi}
\authornote{The first and second authors have the same contribution for the present research.}
\orcid{0000-0002-8118-0802}
\affiliation{
  \institution{IFRS and PUCRS}
  \country{Brazil}}
\email{roben.lunardi@acad.pucrs.br}

\author{Regio A. Michelin}
\affiliation{
  \institution{CSE, UNSW}
  \country{Brazil} 
}
\email{regio.michelin@unsw.edu.au}

\author{Charles V. New}
\affiliation{
  \institution{UNISC}
  \country{Brazil} 
}
\email{charlesneu@gmail.com}

\author{Avelino F. Zorzo}
\affiliation{
\institution{PUCRS}
\country{Brazil}
}
\email{avelino.zorzo@pucrs.br}

\author{Salil S. Kanhere}
\affiliation{
  \institution{CSE, UNSW}
  \country{Australia}}
\email{salil.kanhere@unsw.edu.au}

\renewcommand{\shortauthors}{Lunardi et al.}

\begin{abstract}
The Internet of Things (IoT) is transforming our physical world into a complex and dynamic system of connected devices on an unprecedented scale. Connecting everyday physical objects is creating new business models, improving processes and reducing costs and risks. 
Recently, blockchain technology has received a lot of attention from the community as a possible solution to overcome security issues in IoT. However, traditional blockchains (such as the ones used in Bitcoin and Ethereum) are not well suited to the resource-constrained nature of IoT devices and also with the large volume of information that is expected to be generated from typical IoT deployments. 
To overcome these issues, several researchers have presented lightweight instances of blockchains tailored for IoT.
%
For example, proposing novel data structures based on blocks with decoupled and appendable data.
However, these researchers did not discuss how the consensus algorithm would impact their solutions, \textit{i.e.}, the decision of which consensus algorithm would be better suited was left as an open issue. 
In this paper, we improved an appendable-block blockchain framework to support different consensus algorithms through a modular design. 
We evaluated the performance of this improved version in different emulated scenarios and studied the impact of varying the number of devices and transactions and employing different consensus algorithms. Even adopting different consensus algorithms, results indicate that the latency to append a new block is less than 161ms (in the more demanding scenario) and the delay for processing a new transaction is less than 7ms, suggesting that our improved version of the appendable-block blockchain is efficient and scalable, and thus well suited for IoT scenarios. 
\end{abstract}




\maketitle
\section{Introduction} \label{sec:intro}

The Internet of Things (IoT) refers to the tight integration of devices that are connected to sense, monitor and control processes encompassing various application domains, such as smart homes and smart cities~\cite{Al-Fuqaha:2015}. The IoT is transforming our physical world into a complex and dynamic system of connected devices on an unprecedented scale. Also, it is expected that the widespread adoption of IoT will increase productivity, safety, efficiency and accuracy in different sectors, such as smart factors, supply chain, and health care~\cite{Sharma:2018}. 

Despite the expected benefits, IoT systems potentially present considerable safety and security risks, as they can be used in critical infrastructures such as energy, smart cities and health care. These systems are often a primary target for cybernetic attacks since it is possible to cause significant damage to critical infrastructure and even human lives. Thereby, IoT brings new challenges of network management, overhead in computation, data management and security requirements that need to be addressed efficiently for the large and sensitive amount of data being produced by an ever increasing number of devices, sensors and systems connected together~\cite{Chaudhary:2018}.

In recent years, several researchers~\cite{Huh:2017}~\cite{Pinno:2017}~\cite{DorriPerCom:2017}~\cite{Boudguiga:2017}~\cite{Lunardi:2018}~\cite{Novo:2018} have proposed different solutions that use the blockchain technology in IoT to solve security issues. Some works propose novel blockchain architectures~\cite{DorriPerCom:2017}~\cite{Boudguiga:2017}, while others propose innovative blockchain data management solutions~\cite{Lunardi:2018}~\cite{Novo:2018}. 
However, few blockchain proposals for IoT present a modular framework that can be adapted in different scenarios or easily changed to support different consensus algorithms. 

Consequently, existing research has not addressed the following key challenges: (\textit{i}) a blockchain solution that provides a fast response (few milliseconds) to insert and retrieve data from multiple devices; (\textit{ii}) investigation into the impact of known blockchain attacks in IoT environments; and, (\textit{iii}) deliberation about consensus algorithms and their impact on the IoT context.

In order to fill this gap, the primary goal of this work, is to propose a modular lightweight blockchain framework that can provide a fast response time for insertion of new information and support different device types of consensus algorithms in the IoT context. 

More, specifically, the paper makes the following key contributions: (\textit{i}) a formalization for a modular lightweight blockchain that can be used in gateway-based IoT architecture; (\textit{ii}) a discussion about main security issues in blockchain and how they impact the proposed blockchain considering an IoT scenario; and, (\textit{iii}) an evaluation of the SpeedyChain using two different consensus algorithms to demonstrate its viability in the IoT context.


\section{Background} \label{sec:background}

Blockchain was introduced by Bitcoin~\cite{Nakamoto2008} to ensure a resilient and collaborative solution and to allow transactions between different peers with non-repudiation and tamper-resistance data~\cite{Tschorsch2016}. In the last few years, several blockchain instances have been proposed~\cite{Muneeb:2016} with different purposes, such as: Domain Name System
~\cite{Wang:2017}, Supply Chain~\cite{Bocek:2017}, Vehicular Networks~\cite{DorriComm:2017}~\cite{Yang:2018}
, and Smart Grids~\cite{Zhumabekuly:2016}
~\cite{Guan:2018}. To be used in different domains, these blockchain solutions can consider the usage of different cryptography algorithms, consensus algorithms, data management and block structures.

In order to help understanding these differences and how they impact a blockchain, Zorzo \textit{et al.} ~\cite{Zorzo:2018} categorized blockchain components into four layers: Communication, Consensus, Data and Application. The ``Communication" layer represents how the nodes in a blockchain communicate and exchange information. This layer  defines the communication protocols, P2P architectures, and network infrastructure used by a blockchain. 

Additionally, the ``Consensus" layer encompasses the mechanisms for validating the candidate blocks before inserting them into the ledger and broadcasting that to other peers. The consensus algorithm is required in IoT context since the network is public, and usually there is no trust among peers. 

The ``Data" layer presents the blockchain information structure. This layer specifies the adopted cryptography algorithms, how data are stored, how the access to these data is performed, and how data are replicated. Additionally, there are some different approaches for the data types that are stored in a blockchain. 

Moreover, there are different ways to use a blockchain. The ``Application" layer defines the APIs for using data from a blockchain. For example, there are different ways to access data~\cite{Boudguiga:2017}, to use coins~\cite{Nakamoto2008}, to generate tokens~\cite{Fenu:2018}, to execute a distributed application~\cite{Wang:2017}, to use an identity management~\cite{Lin:2018}, to execute smart contracts~\cite{Atzei:2017}. 


Li \textit{et al.}~\cite{Li:2017} presented a discussion and possible solutions to use blockchain in IoT. They proposed a solution focusing on the ``Communication" layer \cite{Zorzo:2018} using P2P architecture that uses a mechanism called satellite chains, which use validating peers to share information between these chains. Furthermore, they propose integration with Hyperledger Fabric~\cite{hyperledgerfabric:2016}. However, they do not present evaluation of the performance results, nor security analysis of the proposed solution. Consequently, it is hard to evaluate in which scenario their work could be applied to.

Boudguiga \textit{et al.}~\cite{Boudguiga:2017} focused on the ``Application" layer of the blockchain, employing blockchain to perform access control in the context of IoT. Moreover, they present a discussion about the application of their proposal in different scenarios in which IoT is used, such as Smart Homes, Smart Grids, and Industry 4.0. They also presented an infrastructure based in a Blockchain-as-a-Service (BaaS) that is able o improve the application performance. However, their paper does not present practical experiments to support the evaluation, nor the blockchain data management is considered in the research.

Focusing on the ``Consensus" layer, Feng \textit{et al.} \cite{Feng:2018} proposed an Hierarchical Byzantine Fault Tolerant consensus algorithm in order to solve the scale issues presented by PBFT. The idea consists of clustering nodes and setting a leader for each cluster. Only these leaders will perform the consensus. This approach is similar to what is proposed by gateway-based architectures~\cite{DorriPerCom:2017}\cite{Lunardi:2018}. However, they do not present the evaluated architecture nor present how they implemented their solution.

Focusing on the architecture of the ``Communication" layer, Dorri~\textit{et al.}~\cite{DorriPerCom:2017} propose a solution where overlays control the access to data stored in a blockchain shared among different overlays. In this architecture, an overlay has enough computing power to maintain a blockchain and IoT devices are not exposed to common attacks such as Distributed Denial of Service (DDoS) and Dropping Attack~\cite{Johnson:2014}. 

In a similar architecture, Lunardi \textit{et al.}~\cite{Lunardi:2018} proposed the adoption of gateways (limited hardware, however with enough power to maintain a blockchain). Additionally, they presented a different solution for the ``Data" layer, where they introduced the concept of appendable blocks, \textit{i.e.}, a block can continue to be appended with information after it  has been inserted into the blockchain. Also considering the ``Data" layer, in a different work, Dorri \textit{et al.}~\cite{Dorri:2019} proposed deletion of blocks in the blockchain. These two works~\cite{Lunardi:2018}~\cite{Dorri:2019} can help to reduce the amount of data that is managed by nodes in a blockchain, which is important in environments that produce large amount of data.

Additionally, a framework called SpeedyChain~\cite{Michelin:2018}, presents a blockchain to be used in Smart Cities scenarios. Also, to improve the ``Data" layer, SpeedyChain contains a mechanism to control the amount of information inside a block (using expiration of the public key) and a mechanism to detach the payload from the block.

While existing research presented important improvements in the state of the art, few discussions were presented in relation to the security analysis of the proposed solutions and how the ``Consensus" layer choice impact the performance of a blockchain for IoT. Consequently, in the next sections  we present some advances to fill these gaps.

\section{Security issues regarding consensus algorithms}

In this section, we present a discussion about known attacks that could be performed on blockchains and analyze their impact in an IoT setting. 
In order to analyze these attacks, we classified them using the stack model proposed by Zorzo \textit{et al.}~\cite{Zorzo:2018}, as described in Sec.~\ref{sec:background}, and arranged the model to different threats in Table~\ref{tab:security}. Even though we mention different attacks, regarding this paper, we focus on the main attacks that compromise the consensus layer, \textit{i.e.}, 51\% Attack, Block-withholding, Bribery Attack, Double Spending, Finney Attack, Fork-after-withhold, Selfish Mining, Sybil Attack and Vector76 Attack. We briefly describe those attacks next.  

\begin{table}[h!]
\caption{Most common security issues for blockchains}
\begin{tabular}{|l|l|l|}
\hline
\multicolumn{1}{|c|}{\textbf{Threat}}                                      & \multicolumn{1}{c|}{\textbf{Layer}}                                    & \multicolumn{1}{c|}{\textbf{Cause}}                                                                   \\ \hline
Double Spending                                                            & \begin{tabular}[c]{@{}l@{}}Consensus, Data,\\ Application\end{tabular} & \begin{tabular}[c]{@{}l@{}}Concurrency and delay \\ to insert new transactions\end{tabular}           \\ \hline
Finney Attack                                                              & \begin{tabular}[c]{@{}l@{}}Consensus, Data,\\ Application\end{tabular} & \begin{tabular}[c]{@{}l@{}}Concurrency and \\ consensus algorithm\end{tabular}                        \\ \hline
Vector76 Attack                                                            & \begin{tabular}[c]{@{}l@{}}Consensus, Data,\\ Application\end{tabular} & \begin{tabular}[c]{@{}l@{}}Concurrency, mining process \\ and consensus algorithm\end{tabular}        \\ \hline
51\% Attack                                                                & Consensus                                                              & \begin{tabular}[c]{@{}l@{}}Consensus algorithm based  \\on computing power\end{tabular}     \\ \hline
Selfish Mining                                                             & \begin{tabular}[c]{@{}l@{}}Consensus, \\ Communication\end{tabular}                                                     & Fork decision algorithm                                                                               \\ \hline
Block-withholding                                                          & Consensus                                                              & \begin{tabular}[c]{@{}l@{}}Mining pool reward \\ mechanism\end{tabular}                                                                          \\ \hline
\begin{tabular}[c]{@{}l@{}}Fork-after-withhold\\ (FAW)\end{tabular}        & Consensus                                                              & \begin{tabular}[c]{@{}l@{}}Fork decision and mining \\  pool reward mechanism\end{tabular}             \\ \hline
Bribery Attack                                                             & Consensus                                                              & \begin{tabular}[c]{@{}l@{}}Consensus and fork \\ decision algorithm\end{tabular}                      \\ \hline
Deanonymization                                                            & \begin{tabular}[c]{@{}l@{}}Communication, \\Data\end{tabular}                                                          & \begin{tabular}[c]{@{}l@{}}P2P connections and \\ public key reuse\end{tabular}                       \\ \hline
DDoS Attack                                                                & Communication                                                                & Consume target resources                                                                              \\ \hline
\begin{tabular}[c]{@{}l@{}}Transaction \\ Malleability\end{tabular}        & Data                                                                   & \begin{tabular}[c]{@{}l@{}}Bitcoin blockchain  \\transaction id usage\end{tabular}                    \\ \hline
Sybil Attack                                                               & \begin{tabular}[c]{@{}l@{}}Consensus, \\Communication \end{tabular}                                                    & \begin{tabular}[c]{@{}l@{}}P2P network and the ability \\ to create multiple identities\end{tabular} \\ \hline
Eclipse Attack                                                             & Communication                                                                & Network isolation                                                                                     \\ \hline
\begin{tabular}[c]{@{}l@{}}Smart Contracts \\ Vulnerabilities\end{tabular} & Application                                                            & \begin{tabular}[c]{@{}l@{}}Bad programming practices\\   and Smart contract errors\end{tabular}        \\ \hline
\end{tabular}
\label{tab:security}
\end{table}

Double Spending, Finney, Vector 76\%, and Transaction Malleability attacks are aimed at spending coins in multiple transactions. In \textbf{Double Spending attack}~\cite{Conti:2018}, a malicious user sends multiple transactions to reachable peers in order to spend the same coin more than once. Alternatively, \textbf{Finney attack}~\cite{Conti:2018} consists of a dishonest miner holding a pre-mined block, and spending the same coin that is used in a transaction of the pre-mined block. Combining these two attacks, \textbf{Vector 76\% attack}~\cite{Conti:2018} consists of requesting to withdraw the value of a transaction that was confirmed and sending the same value to another transaction, exploring the fork resolution algorithm (generating conflict in the longest chain). 

Many proposals that adopt blockchain in IoT scenarios~\cite{Li:2017}~\cite{Boudguiga:2017} \cite{DorriPerCom:2017}~\cite{Lunardi:2018}~\cite{Michelin:2018} do not use cryptocurrencies. Consequently, Double spending, Finney, and Vector 76 
attacks are not attractive for malicious users. For example, in the case of SpeedyChain~\cite{Michelin:2018}, an appendable-block blockchain, these attacks do not represent a threat as sending multiple transactions with the same timestamp, signature, and information will be discarded in case of collision or in case of incorrect order, the transaction will be discarded. 

There are different attacks that explores vulnerabilities in the mining mechanism of Proof-of-Work (consensus algorithm), such as 51\%, Selfish Mining, Block-Withholding, Fork-After-Withholding, and Bribery attacks. The \textbf{51\% attack} consists of a malicious user controlling more than 50\% of network processing power, thus this user could rewrite the blockchain blocks and define the blockchain behavior~\cite{Gervais:2016}. Similarly, \textbf{Selfish Mining} attack consists of a malicious user (or a pool), keeping own mined blocks private until its chain reach a longer length than the main blockchain. As per the fork rule, the attacker chain will now become the main chain~\cite{Eyal:2014}. \textbf{Block-Withholding} happens when a malicious miner - which is participating in a mining pool - finds a valid hash value and sends it directly to the blockchain network, thus avoiding division of the reward for mining the block~\cite{Bag:2017}. Similarly, in \textbf{Fork-After-Withhold (FAW)} a malicious miner holds the block until another miner (from the same pool) identifies a block. Then, the malicious miner sends its block, forcing the pool to generate a fork (this block could be sent do multiples pool in order to increase its reward)~\cite{Kwon:2017}. \textbf{Bribery attack}~\cite{Bonneau:2016} consists of a malicious user exploring the mining power of different nodes (through financial incentives) to include conflicting transactions in the blockchain (\textit{e.g.}, can be used to force a Double Spending).
\textbf{Sybil attack} relies on a malicious node assuming multiples identities in the network with the ultimate goal of influencing the network~\cite{Douceur:2002}. The \textbf{Eclipse attack} consists of a malicious user aiming to monopolize the incoming and outgoing connections of a victim, thus isolating the victim from the main blockchain network~\cite{Heilman:2015}. 

51\%, Selfish Mining, Block-Withholding, FAW and Bribery attacks are based on strategies adopted by PoW (Proof-of-Work) consensus algorithms. Consequently, choosing a solution for IoT that use a different consensus algorithm (\textit{e.g.}, PBFT) can help to avoid these kind of attacks. A key aspect to be considered is related to the hardware constraints in IoT devices, such as computing power, memory, and storage. 
In order to solve these issues, we proposed (see Section~\ref{section:consensus}) and evaluated (see Section~\ref{sec:discussion}) in this paper the use of two different consensus algorithms for IoT environments.

\section{Appendable-block blockchain in IoT} \label{sec:speedychain}


In this section, we present the fundamental concepts of a blockchain architecture that underpins our proposed framework. 

\begin{figure}[h!]
    \centering
    \includegraphics[width=0.48\textwidth]{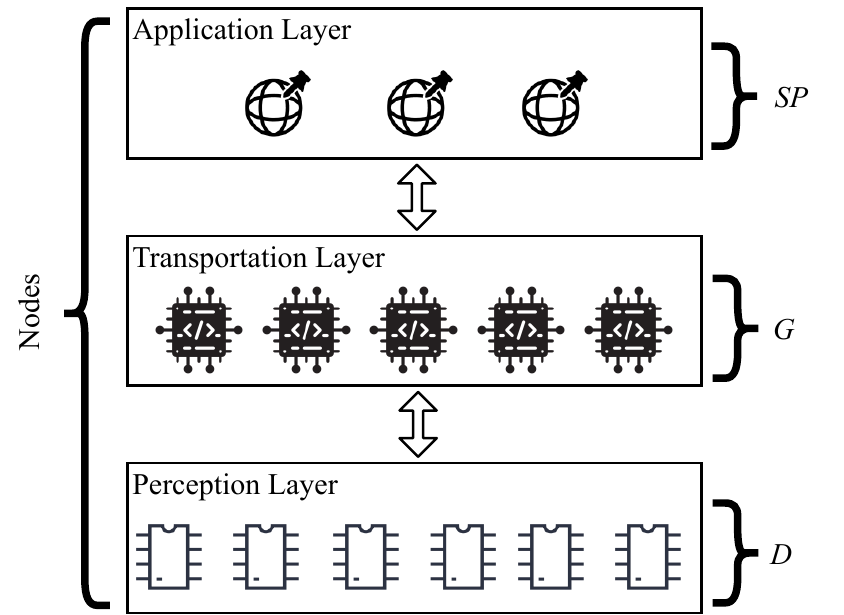}
    \caption{Main IoT nodes}
    \label{fig:architecture}
\end{figure}

The proposed framework was designed using a layer-based IoT architecture~\cite{Jing2014} - similar to that is adopted in Lunardi \textit{et al.}~\cite{Lunardi:2018}, Dorri \textit{et al.}~\cite{DorriPerCom:2017} and Michelin \textit{et al.}~\cite{Michelin:2018} - that is composed by: (\textit{i}) devices (\textit{D} in Fig.~\ref{fig:architecture})  in the Perception Layer; (\textit{ii}) Gateways (\textit{G}) in the Transportation Layer; and (\textit{iii}) Service Providers (\textit{SP}) in the Application Layer. Therefore, each device can produce information and send to the gateways to append data to its own block. Consequently, devices can keep producing and appending information into blockchain independently to the other devices operations. Service Providers can access the information from a device (that it is stored in the blockchain) through the gateways.

It is important to note that this work focuses on the blockchain that is maintained in the Gateway Level presented in the IoT architecture (Fig. \ref{fig:architecture}). Also, it is important to note that this work uses concepts that were presented in other works~\cite{Lunardi:2018}~\cite{Michelin:2018}, but adapted them for a more dynamic scenario using a modular blockchain. Consequently, consensus algorithms can be used based on each IoT requirement.  Moreover, the proposed solution was designed to maintain the integrity and availability of the data collected from different sensors/devices for both audition and control (by an application or based on predefined rules), based on predefined policies for each device (in the Device Level). The proposed solution provides an API to applications for internal (\textit{e.g.}: logs, alerts, and logistics) or external usage (\textit{e.g.} providing APIs for partners applications).

\subsection{Architecture}\label{sec:achitecture}

Let \textit{N} = \{\textit{N\textsubscript{1}}, ..., \textit{N\textsubscript{n}}\}
be the set of \textit{n} nodes in the system with public-private key pairs (\textit{NPK\textit{\textsubscript{i}}}, \textit{NSK\textit{\textsubscript{i}}}). Also, consider that these nodes can have different roles in the architecture. Consequently, this system  is composed by \textit{d} devices, where \textit{D} = \{\textit{D\textsubscript{1}}, ..., \textit{D\textsubscript{d}}\}, that usually produce information and could be controlled remotely; \textit{g} gateways, where \textit{G} = \{\textit{G\textsubscript{1}}, ..., \textit{G\textsubscript{g}}\}, that manage the access to information in a blockchain; not limited to this, different kind of nodes are supported such as \textit{s} service providers \textit{SP} = \{\textit{SP\textsubscript{1}}, ..., \textit{SP\textsubscript{s}}\}. Therefore, \textit{N\textsubscript{i}} = \textit{\{D, G, SP\}}. Assume that all nodes in N can use the same cryptography algorithms. Moreover, every \textit{NPK\textit{\textsubscript{i}}} should be different and accessible by any participant in this system. Also, assume that a key pair (public and secret keys) from a device will be represented as (\textit{DPK\textit{\textsubscript{j}}}, \textit{DSK\textit{\textsubscript{j}}}) and a key pair from gateway will be represented as (\textit{GPK\textit{\textsubscript{h}}}, \textit{GSK\textit{\textsubscript{h}}}). Consider that each device in \textit{D} (Devices Level) should be connected to a gateway in \textit{G} (Gateway Level) through different (wired or wireless) network devices (Network Level). Additionally, the gateways are responsible to manage the device access and provide an API that allows to manage the blockchain.




\subsection{Blockchain Definition}\label{sec:blockchaindefinition}

\begin{figure}[h!]
    \centering
    \includegraphics[width=0.82\textwidth]{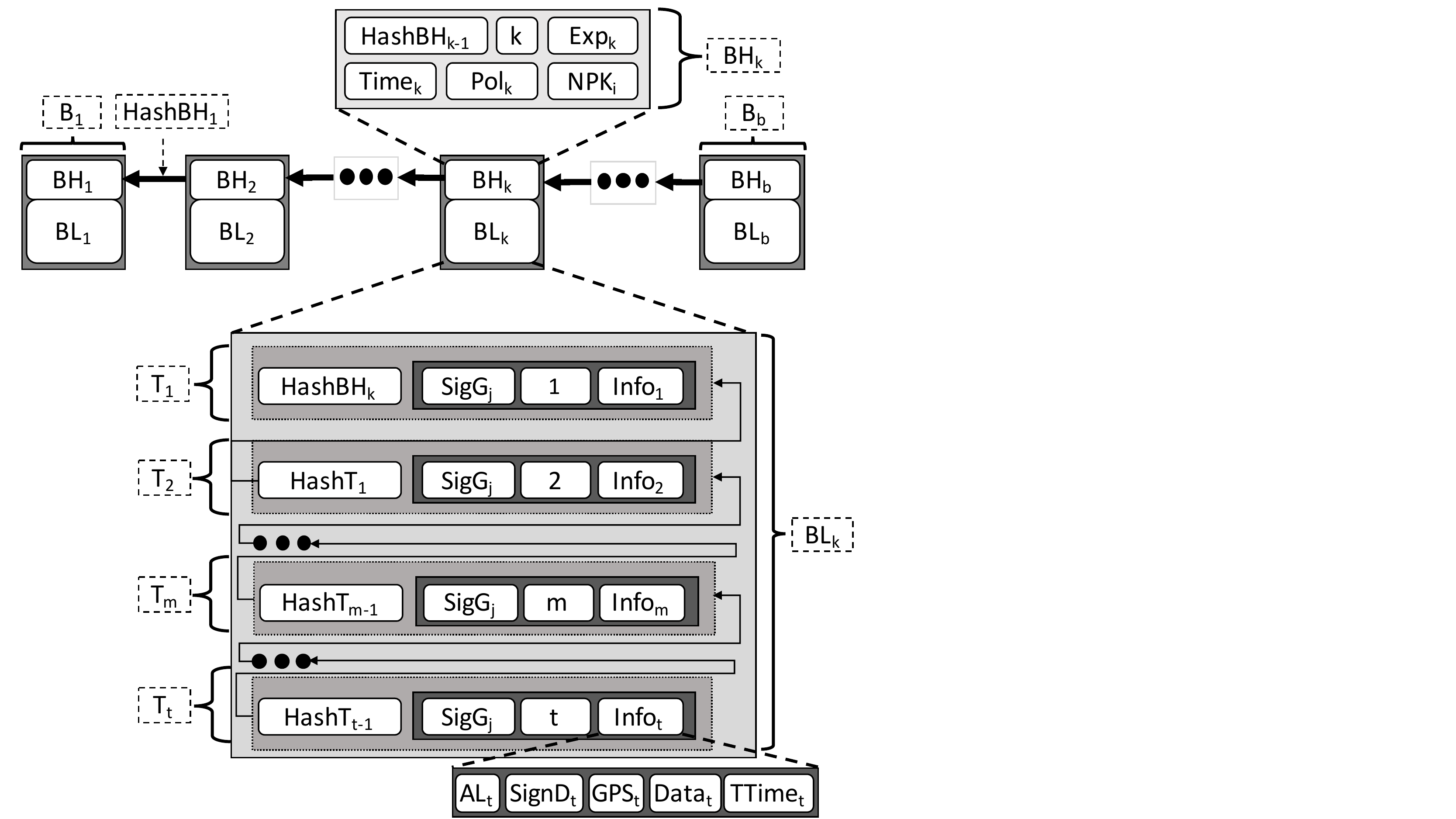}
    \caption{Main blockchain components.}
    \label{fig:blockchain}
\end{figure}

Based on the IoT architecture presented in Fig.~\ref{fig:architecture}, the blockchain will be maintained by gateways in \textit{G} (Gateway Level in Fig.~\ref{fig:architecture}).
To ensure that every participant can access any \textit{NPK\textit{\textsubscript{i}}} (\textit{e.g.}, \textit{DPK\textit{\textsubscript{j}}} or \textit{GPK\textit{\textsubscript{h}}}) and information stored in a Gateway was not tampered with, let a blockchain \textit{B = \{\textit{B\textsubscript{1}}, ..., \textit{B\textsubscript{b}}\}} be a set of \textit{b} blocks. 
Each \textit{B\textit{\textsubscript{k}}} has a pair of different information (\textit{BH{\textsubscript{k}}, BL{\textsubscript{k}}}), where \textit{BH{\textsubscript{k}}} is responsible to maintain the block header of \textit{B{\textsubscript{k}}} and the \textit{BL{\textsubscript{k}}} stores the block ledger, \textit{i.e.}, the set of transactions of \textit{B{\textsubscript{k}}} as shown in details in Fig.~\ref{fig:blockchain}.

Therefore, \textit{BH\textsubscript{k}} is composed by (\textit{HashBH\textsubscript{k-1}, k, Time\textsubscript{k}, Exp\textsubscript{k}, Pol\textsubscript{k}, NPK\textsubscript{i}}), where 
\[HashBH\textsubscript{k-1} = \left\{
  \begin{array}{lr}
    0 & \text{, when } k  = 1\\
    \text{hash digest of \textit{BH\textsubscript{k-1}}} & \text{, when } k \ge 2
  \end{array}
\right.
\]
where hash digest is obtained through a hash function, \textit{i.e.}, \textit{HashBH\textsubscript{k-1}} contains the hash digest of previous block header (or zero when it is the first block); \textit{k} is equal to the index of the block \textit{B\textsubscript{k}} in the blockchain; \textit{Time\textsubscript{k}} is the timestamp from when the block was generated; \textit{Exp\textsubscript{k}} presents the threshold time to insert a new transaction in its block ledger, for example, after this time a device should create a new key pair (\textit{NPK}, \textit{NSK}) and create a new block; \textit{Pol\textsubscript{k}} presents the access policy that the device has to attend; and \textit{NPK\textsubscript{j}} is the node public key. It is important to mention that every node - independent of its type - should have a block in B, composed of at least a block header, and every \textit{NPK} should be available in the blockchain.

Let \textit{BL\textsubscript{k}}=\{\textit{T\textsubscript{1}, ..., T\textsubscript{t}\}} be the set of \textit{t} transactions on the block ledger of the block \textit{B\textsubscript{\textit{k}}}.\textit{ T\textsubscript{m}} is composed by (\textit{HashT\textsubscript{m-1}, m, SigG\textsubscript{m}, Info\textsubscript{m}}), where 

\[HashT\textsubscript{m-1} = \left\{
  \begin{array}{lr}
    \text{hash digest of } BH\textsubscript{k} & \text{, when } m  = 1\\
    \text{hash digest of } \textit{T\textsubscript{m-1}} & \text{, when } m \ge 2
  \end{array}
\right.
\]
where the \textit{HashT\textsubscript{m-1}} contains the hash of the previous transaction (or the hash of its block header when it is the first transaction of the block ledger); \textit{m} is equal to the index of the transaction T\textsubscript{m} in the block ledger \textit{BL\textsubscript{k}}; \textit{SigG\textsubscript{m}} represents the result of the cryptography using the \textit{GPK\textit{\textsubscript{h}}} to sign \textit{Info\textsubscript{m}}. 

The \textit{Info\textsubscript{m}} can be different for each type of node. Devices provide a set of information (\textit{SigD\textsubscript{m}, AL\textsubscript{m},GPS\textsubscript{m}, Data\textsubscript{m}, TTime\textsubscript{m}}), where \textit{AL\textsubscript{m}} is the access level required to access the information from outside of the blockchain that is defined by the device \textit{D\textsubscript{j}}, while the \textit{SigD\textsubscript{m}} represents the signature of (\textit{AL\textsubscript{m}}, \textit{GPS\textsubscript{m}}, \textit{Data\textsubscript{m}}, and \textit{TTime\textsubscript{m}}) using \textit{DPK\textsubscript{j}}, where \textit{GPS\textsubscript{m}} represents the global position of the device (when it is available), while Data\textsubscript{m} is the data collected/set from/to device \textit{D\textsubscript{j}} and \textit{TTime\textsubscript{m}} is the timestamp when the \textit{Data\textsubscript{m}} was generated/set. It is important to note that \textit{Data\textsubscript{m}} could be formatted differently depending on the device. For example, it could store a single read of a sensor (an integer type) or a set of information, encrypted or not, depending on the configuration established in the API level. 

\subsection{Main Operations}\label{section:mainoperations}

The main operations that can be performed in the proposed blockchain are: appending blocks, appending transactions, key update and consensus algorithm. They are detailed in the next subsections.

\subsubsection{Appending blocks}\label{section:appendblock}

Insertion of a new block\textit{ B\textsubscript{k}} in blockchain \textit{B} is started by a gateway (present in Gateway level) with the objective to include a new node (\textit{N\textsubscript{i}}) public key (NPK\textsubscript{i}). This algorithm is performed every time that a node \textit{N\textsubscript{i}} requests a connection and its Public Key (\textit{NPK\textsubscript{i}}) is not present in the blockchain (line 1 in Algorithm~\ref{alg:appendblock}). 

After verifying that a NPK\textsubscript{i} is not present in the blockchain, the gateway should send this new public key to perform a consensus to insert the new block (line 2). It is important to note that the consensus is performed by a Leader elected in the blockchain (see details in Sec.~\ref{section:consensus}).

 \begin{algorithm}[H]
 \caption{Insertion of new blocks in the blockchain}
 \label{alg:appendblock}
\begin{algorithmic}[1]
\REQUIRE Connection request and requester $NPK\textsubscript{i}$
\IF {$NPK\textsubscript{i}$ is not present in any BH\textsubscript{j}}
\STATE {\textbf{sendBlockToConsensus}($NPK\textsubscript{i}$)}
\ENDIF 

\end{algorithmic}
\end{algorithm}

\subsubsection{Appending transaction}\label{section:appendtransaction}

Every time a node \textit{N\textsubscript{i}} produces new information \textit{Info\textsubscript{m}} to be inserted in the blockchain, it has to communicate to a gateway to append the transaction to its block ledger \textit{BL\textsubscript{i}}. This operation is performed only if the node public key (\textit{NPK\textsubscript{i}}) is present in a block header \textit{BH\textsubscript{i}} from blockchain \textit{B} (line 1 in Algorithm~\ref{alg:appendtrans}). When, a gateway receives a new information \textit{Info\textsubscript{m}}, the digital signature \textit{SigD\textsubscript{m}} present in \textit{Info\textsubscript{m}} should be validated (lines 2 and 3) using the public key \textit{NPK\textsubscript{i}}.

After the validation of the signature, the gateway performs the encapsulation of the new transaction, setting: the hash of the previous transaction \textit{HashT\textsubscript{m-1}} (line 6), the index of the transaction (based on the last transaction) \textit{m} (line 7), and the digital signature from the gateway that is processing the transaction \textit{SigG\textsubscript{m}} (line 8) using its secret key \textit{GSK\textsubscript{h}}. 

After that, the gateway creates the new transaction \textit{T\textsubscript{m}} (line 9), and the transaction can be broadcast to the other gateways (line 10).  

 \begin{algorithm}[H]
 \caption{Appending new transactions into the block ledger}
 \label{alg:appendtrans}
\begin{algorithmic}[1]
\REQUIRE {$Info\textsubscript{m}$ and device $NPK\textsubscript{i}$}
\IF {$NPK\textsubscript{i}$ is present in any $BH\textsubscript{j}$}
\STATE {$result \gets$ \textbf{verifySign}($NPK\textsubscript{i},Info\textsubscript{m}$)}
\IF {$result$ is \TRUE}
\STATE {$b\gets$ \textbf{blockIndex}$(B,NPK\textsubscript{i})$}
\STATE {$t\gets$ \textbf{lastTransaction}($BL\textsubscript{b}$)}
\STATE {$HashT\textsubscript{m-1}\gets$ \textbf{hash}($T\textsubscript{t})$}
\STATE $m\gets t+1 $
\STATE {$SigG\textsubscript{m}\gets$ \textbf{sign}($GSK\textsubscript{h},Info\textsubscript{m}$)}
\STATE {$T\textsubscript{m}\gets$ \{$HashT\textsubscript{m-1},m,SigG\textsubscript{m},Info\textsubscript{m}$\}}
    \STATE {\textbf{broadcast}($T\textsubscript{m}, BH\textsubscript{b}$)}
\ENDIF 
\ENDIF
\end{algorithmic}
\end{algorithm}

\subsubsection{Key Update}\label{section:KeyUpdate}

Anytime that a gateway receives a transaction with its timestamp \textit{TTime\textsubscript{m}} with a higher value than the expiration time present in the origin node \textit{N\textsubscript{i}} expiration time \textit{Exp\textsubscript{k}} the gateway will execute the key update algorithm (Algorithm~\ref{alg:keyupdate}). Also, the node \textit{N\textsubscript{i}} can send to the gateway a request to update its public key \textit{NPK\textsubscript{i}'}. 

In both situations, a gateway will request the node \textit{N\textsubscript{i}} its new public key \textit{NPK\textsubscript{i}'} (line 1 in the Algorithm~\ref{alg:keyupdate}). After the key validation (\textit{e.g.}, if the key is not already in the blockchain), the gateway will append a new block into the blockchain with the new \textit{NPK\textsubscript{i}'} from node \textit{N\textsubscript{i}} (line 3). 

In order to append a new block, a gateway will use Algorithm~\ref{alg:appendblock} presented previously. Consequently, each node will receive a new block with the new public key \textit{NPK\textsubscript{i}'} of the node \textit{N\textsubscript{i}}.

 \begin{algorithm}[H]
 \caption{Algorithm for key update}
  \label{alg:keyupdate}
\begin{algorithmic}[1]
\REQUIRE {$TTime\textsubscript{m} \ge Exp\textsubscript{k}$ or requested by node $N\textsubscript{i}$}
\STATE {$NPK\textsubscript{i}' \gets $ \textbf{requestNewKey}($NPK\textsubscript{i}$)}
\IF {$NPK\textsubscript{i}'$ is valid}
\STATE {\textbf{appendBlock}($NPK\textsubscript{i}'$)} \COMMENT{see Algorithm~\ref{alg:appendblock}}

\ENDIF 
\end{algorithmic}
\end{algorithm}

\subsubsection{Consensus}\label{section:consensus}

Usually, a blockchain was designed to allow the adoption of different consensus algorithms. Before discussing different consensus algorithms, first we need to present what is a valid block or transaction. For a transaction to be considered valid, it should have a \textit{NPK\textsubscript{i}} that is already in the blockchain, a valid signature (based on the data transmitted and \textit{NPK\textsubscript{i}}), and a \textit{TTime\textsubscript{m}} lower than its  \textit{Exp\textsubscript{k}} (present in the block header) to ensure that no transactions are inserted in an expired block. Moreover, to ensure that a block header is valid: (\textit{i}) the gateways should agree that a new node \textit{NPK\textsubscript{i}} can be part of the blockchain \textit{B}; (\textit{ii}) the access policy \textit{Pol\textsubscript{k}} for this node \textit{NPK\textsubscript{i}} should be defined; (\textit{iii}) the \textit{Exp\textsubscript{k}} should be calculated to avoid a large block in size. In this work we assume that this validation is performed by the gateways through predefined rules. 

Currently, there are different consensus algorithms used by blockchains, such as: Proof-of-Work (PoW), Proof-of-Stake (PoS),  Byzantine Fault-Tolerance (PBFT), Federated Byzantine Agreement (FBA) or delegated Byzantine Fault-Tolerance (dBFT). Furthermore, it is not possible to define a single solution that will perform better than others for any scenario. 

Two different consensus algorithms are proposed, but not limited to them : (\textit{i}) validation based on a specific number of witness, where every block should be signed by at least a predefined number of witness; and (\textit{ii}) adapted PBFT algorithm, where more than 2/3 of the active gateways should validate and sign the block. Both consensus algorithms could be summarized in Algorithm~\ref{alg:consensus}.

 \begin{algorithm}[h!]
 \caption{Generic consensus algorithm}
  \label{alg:consensus}
\begin{algorithmic}[1]
\REQUIRE {receive a $NPK\textsubscript{i}$ to perform consensus}
\STATE {$b\gets \textbf{lastIndex}(B)$}
\STATE {$HashBH\textsubscript{k-1}\gets$ \textbf{hash}($BH\textsubscript{b}$)}
\STATE $k\gets b+1 $
\STATE {$Time\textsubscript{k}\gets$ \textbf{getTime()}}
\STATE {$Exp\textsubscript{k} \gets$ \textbf{defineExp()}}
\STATE {$Pol\textsubscript{k} \gets$ \textbf{setPolicy()} }
\STATE {$BH\textsubscript{k}\gets$ \{$HashBH\textsubscript{k-1},k,Time\textsubscript{k},Exp\textsubscript{k},Pol\textsubscript{k},NPK\textsubscript{i}$\}}
\STATE {$consensusResponses\gets$ \textbf{performConsensus}($BH\textsubscript{k}$)}
\IF {$consensusResponses > minimumResponses$}
    \STATE {\textbf{broadcast}($BH\textsubscript{k}$)}
\ENDIF 

\end{algorithmic}
\end{algorithm}

In order to encapsulate the new block~\textit{B\textsubscript{k}}, every information from the block header~\textit{BH\textsubscript{k}} is set, such as the hash of the previous block header \textit{BH\textsubscript{b}} (line 2), block index \textit{k} (line 3), the timestamp using the time of block creation \textit{Time\textsubscript{k}} (line 4), an expiration time \textit{Exp\textsubscript{k}} to control the validity of the block (line 5), and the access policy \textit{Pol\textsubscript{k}} that the new node is submitted to (line 6 in Algorithm~\ref{alg:consensus}). 
It is important to note that both \textit{Exp\textsubscript{k}} and \textit{Pol\textsubscript{k}} are defined at API level. After the block header is created, the consensus is performed (line 7). It is important to note that the consensus is performed only by gateway nodes. After the consensus is performed and it receives more than the minimum responses for each consensus algorithm, the new block is broadcast to the peers (line 10).

We presented a simplified version of consensus algorithm to represent both PBFT and Witness-based consensus algorithms. However, we intend to evaluate other consensus algorithms in a future work, such as dBFT and FBA. 

Next section presents a discussion about overhead introduced by consensus algorithms in the improved appendable-block blockchain.

\section{Performance Evaluation} \label{sec:discussion}

In order to evaluate the performance of the proposed blockchain in IoT scenarios, the CORE emulator platform~\cite{CORE2008} was used. 
The evaluation was run on a VMware Fusion 8.5.10 with 6 processors and \textit{12GB} of \textit{RAM} on an Intel \textit{i7@2.8Ghz} and \textit{16GB} of \textit{RAM}. We performed the evaluation using 10 gateways, where each gateway runs in a container based-virtualized machine; in 9 different scenarios (as presented in Table~\ref{tab:performance}) using 100, 500 and 1000 devices connected through theses gateways (10, 50 and 100 per gateway) and 100, 500 and 1,000 transactions per device (\textit{e.g.}, 1,000,000  transactions in Scenario I). All times presented in Table~\ref{tab:performance} represent the median time considering the whole execution in all gateways.  

\begin{table*}[h!]
\caption{Performance Evaluation}
\begin{tabular}{l|l|l|l|l|l|l|l|l|l|}
\cline{2-10}
                                                                                                                     & A      & B      & C       & D      & E       & F       & G       & H       & I         \\ \hline
\multicolumn{1}{|l|}{Devices per Gw}                                                                                 & 10     & 10     & 10      & 50     & 50      & 50      & 100     & 100     & 100       \\ \hline
\multicolumn{1}{|l|}{Transactions per Device}                                                                        & 100    & 500    & 1,000   & 100    & 500     & 1,000   & 100     & 500     & 1,000     \\ \hline
\multicolumn{1}{|l|}{Total of Devices' Blocks}                                                                                   & 100    & 100    & 100     & 500    & 500     & 500     & 1,000   & 1,000   & 1,000     \\ \hline
\multicolumn{1}{|l|}{Total of Transactions}                                                                             & 10,000 & 50,000 & 100,000 & 50,000 & 250,000 & 500,000 & 100,000 & 500,000 & 1,000,000 \\ \hline
\multicolumn{1}{|l|}{Block Consensus (Witness)}                                                                   & 58.20ms  & 64.01ms  & 65.25ms   & 64.51ms  & 71.02ms   & 71.73ms   & 69.13ms   & 72.47ms   & 79.22ms     \\ \hline
\multicolumn{1}{|l|}{Block Consensus (PBFT)}                                                                      & 102.82ms & 119.53ms & 121.68ms  & 121.98ms & 126.56ms  & 132.37ms  & 129.14ms  & 136.86ms  & 160.35ms    \\ \hline
\multicolumn{1}{|l|}{\begin{tabular}[c]{@{}l@{}}Add Block in Leader (Wit.) \end{tabular}}                    & 3.72ms   & 3.56ms   & 4.42ms    & 4.66ms   & 4.82ms    & 5.81ms    & 5.33ms    & 5.95ms    & 6.28ms      \\ \hline
\multicolumn{1}{|l|}{\begin{tabular}[c]{@{}l@{}}Add Block in Leader (PBFT)\end{tabular}}                    & 3.40ms   & 4.45ms   & 5.16ms    & 4.21ms   & 4.87ms    & 5.88ms    & 5.29ms    & 5.93ms    & 6.52ms      \\ \hline
\multicolumn{1}{|l|}{\begin{tabular}[c]{@{}l@{}}Update Blockchain w/ Block (Wit.)\end{tabular}}       & 0.22ms   & 0.22ms   & 0.23ms    & 0.22ms   & 0.23ms    & 0.23ms    & 0.23ms    & 0.24ms    & 0.25ms      \\ \hline
\multicolumn{1}{|l|}{\begin{tabular}[c]{@{}l@{}}Update Blockchain w/ Block (PBFT)\end{tabular}}      & 0.22ms   & 0.22ms   & 0.23ms    & 0.23ms   & 0.23ms    & 0.26ms    & 0.24ms    & 0.24ms    & 0.27ms      \\ \hline
\multicolumn{1}{|l|}{\begin{tabular}[c]{@{}l@{}}Append Transaction in Gw. (Wit.)\end{tabular}}      & 2.66ms   & 2.82ms   & 2.91ms    & 3.24ms   & 3.49ms    & 3.54ms    & 3.89ms    & 4.29ms    & 4.28ms      \\ \hline
\multicolumn{1}{|l|}{\begin{tabular}[c]{@{}l@{}}Append Transaction in Gw. (PBFT)\end{tabular}}      & 2.69ms   & 2.80ms   & 2.90ms    & 3.30ms   & 3.46ms    & 4.00ms    & 3.96ms    & 4.16ms    & 4.55ms      \\ \hline
\multicolumn{1}{|l|}{\begin{tabular}[c]{@{}l@{}}Update Blockchain w/ Trans. (Wit.)\end{tabular}} & 0.94ms   & 1.18ms   & 1.48ms    & 1.30ms   & 1.58ms    & 1.89ms    & 1.73ms    & 2.11ms    & 2.33ms      \\ \hline
\multicolumn{1}{|l|}{\begin{tabular}[c]{@{}l@{}}Update Blockchain w/ Trans. (PBFT)\end{tabular}} & 0.94ms   & 1.17ms   & 1.47ms    & 1.31ms   & 1.55ms    & 2.03ms    & 1.73ms    & 2.03ms    & 2.39ms      \\ \hline
\end{tabular}
\label{tab:performance}
\end{table*}

The Witness-based consensus was used as a baseline in terms of time to append blocks and information. As expected, it can be observed in Table.~\ref{tab:performance} that varying the consensus algorithm has impact in the performance in the task to achieve consensus on inserting a block (used to insert block header with public key of each device). For example, in Scenario A, witness-based consensus takes 58.20ms to achieve the consensus against 102.82ms using PBFT and in Scenario I (scenario with highest number of devices and transactions), witness-based consensus takes 72.47ms against 160.35ms using PBFT (more than twice the time). However, witness-based consensus is more likely to be affected by different attacks (\textit{e.g.}, Eclipse and Sybil attacks) in comparison to PBFT. 

In the other blockchain operations - for instance, time to add a new block in the leader gateway (gateway that started the consensus), as well as the time to update the blockchain, to append a new transaction in a gateway (where devices are connected to) and to update the blockchain with the new transaction - presented few or no impact using both consensus algorithms. However, the number of transactions and nodes influenced in the processing time to append a transaction in the most demanding scenario (Scenario I) takes less than 7ms to both append the transaction (4.28ms in Witness-based and 4.55ms in PBFT) and to update a new transaction in the other gateways (2.33ms in Witness-based and 2.39ms in PBFT).

\begin{figure}[h!]
    \centering
    \includegraphics[width=0.485\textwidth]{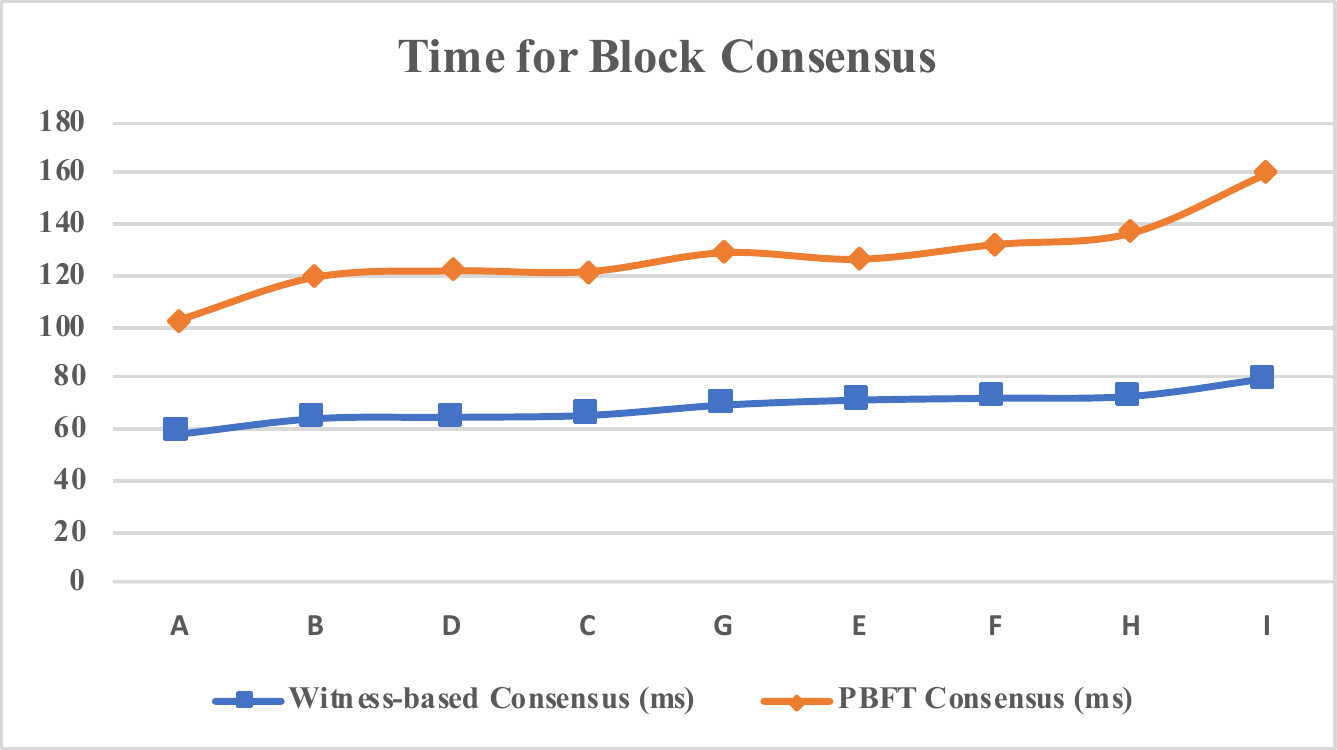}
    \caption{Time for block consensus}
    \label{fig:results_opt6}
\end{figure}

Additionally, it can be observed that growing the number of transactions (overload of processing in gateways) has more impact than the number of devices that a gateway is handling. For example, scenario D has half of transactions and 5 times more nodes than C, but takes almost the same time to reach the consensus for a block. Differently, scenario F has half of nodes and 5 times more transactions than scenario G, resulting in F spending around 3\% more time to achieve the consensus than G. Fig.~\ref{fig:results_opt6} presents a comparison of the time to achieve consensus of a block in different scenarios.

As a  comparison, Bitcoin network has around 10,000~\cite{bitonodes:2018} active nodes in a 24-hour slice, consequently, the experiment in Scenario I represents approximately 10\% of the Bitcoin network. 
As a comparison, Bitcoin has more than 150,000 confirmed transactions per day~\cite{Conti:2018} with a peak of 490,644 confirmed transaction in a day~\cite{transactionsperday:2018}, which means that the evaluation in Scenario I, at least represents more than twice the transactions in the Bitcoin blockchain in a day. A more effective comparison could be made with IOTA~\cite{iota} - a blockchain developed for IoT - which has around 8.7 transactions per second~\cite{IOTAtransactions:2018}. 
This means around 750,000 transactions processed in a day (around 75\% of the transactions processed in Scenario I). Also, it represents that IOTA transaction processing time is around 115ms. Consequently, the transactions processing time in our solution represents less than 6\% of the time that is spent in IOTA - ~115ms in IOTA and ~7ms in our solution (4.55ms to append a transaction in a gateway summed with 2.39ms to update the entire blockchain using PBFT).

The evaluation performed in this paper presented good results in the emulated IoT scenarios with different number of devices and transactions. It is important to note that the code that implements the proposed blockchain was developed using the Python programming language and a set of libraries. The code is available at GitHub and could be used to replicate the experiments (details omitted to the double-blind review). In a future work, we intend to evaluate the solution in a real IoT scenario, composed by different hardware with different number of gateways and mission critical devices. 

\section{Threats to Validity}\label{sec:threats}

In this section, we describe the threats to the validity of the results presented in the evaluation. The first threat is related to hardware capability. In this work, we did not present an evaluation with real IoT devices. However, we used the same cryptography algorithms and methods than those that were adopted by Lunardi \textit{et al.}~\cite{Lunardi:2018} in their experiments (using real hardware). Consequently, devices using IoT hardware should be capable to execute the same operations, but probably with a different performance.

The second threat is related to the architecture adopted and possible malicious gateways performing an Eclipse attack against some devices. Although we assumed that a device can connect to another gateway, we did not discuss this situation in this paper, and therefore, at the moment, our solution is susceptible to an Eclipse attack.  This specific threat should be better addressed in future work. 

Another threat that can affect the evaluation is the mobility of nodes. We did not consider in our evaluation the problems that a mobile device or gateway can produce. Hence, this threat can be further discussed in future work.

\section{Conclusion}\label{sec:conclusion}

Industry 4.0 is increasing the number of devices and the intelligence in these devices. This leads to the need for a data handling that is able to run in a decentralized scenario and at the same time to keep its integrity and resilience with a very fast response time (milliseconds), using consensus algorithms that can be adapted for IoT scenario. To fulfill this need the proposed blockchain presents promising results (using both a simplified Witness-based and PBFT consensus) based on the evaluation performed in Sec.~\ref{sec:discussion}. 

This paper also presented a modular definition of the proposed blockchain and its main operations, which is the capability to handle transactions, appending them to an existing block, and still keep data integrity. Due to this feature, the time to add transactions in a block is kept in a few milliseconds. In comparison to blockchain such as, the ones used in Bitcoin and IOTA, in the proposed blockchain time to include a transaction is considerable lower. Also, due to the proposed modularization, the evaluation was performed using two different consensus algorithms in 9 different scenarios.

A security analysis was conducted in order to discuss the most common attacks that could affect a blockchain consensus layer. It was observed that malicious gateways could interfere or delay the transaction inclusion in the blockchain. Thus, leading to an open issue, \textit{i.e.}, to improve and mitigate attacks such as Eclipse and Sybil.

As future work, we intend to  scale the present scenario  varying the number of gateways that are available. As pointed in the evaluation section, depending on  the gateway processing load, the transaction processing time increases. A further discussion should be performed considering different consensus algorithms.

\begin{acks}
This paper was achieved in cooperation with HP Brasil using incentives of Brazilian Informatics Law (Law n 8.248 of 1991). This study was financed in part by the Coordena\c{c}\~ao de Aperfei\c{c}oamento de Pessoal de N\'ivel Superior - Brasil (CAPES) - Finance Code 001. Also, we thank to Australian Academy of Science for the Australia-Americas PhD Research Internship Program.
\end{acks}

\bibliographystyle{ACM-Reference-Format}
\bibliography{sample-base}

\appendix

\end{document}